\documentclass[aps,prl,twocolumn,showpacs,preprintnumbers,amsmath,amssymb,groupedaddress]{revtex4}

\usepackage{amscd,color,bm}
\usepackage{graphicx}

\newcommand{\vect}[1]{\mathbf{#1}}
\newcommand{\hvect}[1]{\hat{\mathbf{#1}}}

\newcommand{\betaP}{\beta^{\rm P}}
\newcommand{\betaE}{\beta^{\rm E}}

\newcommand{\F}{\mathcal{F}}
\newcommand{\bx}{\mathbf{x}}

\begin{document}

\title{Bending crystals: Emergence of fractal dislocation structures}

\author{Yong S. Chen, Woosong Choi, Stefanos Papanikolaou, James P. Sethna}

\affiliation{Laboratory of Atomic and Solid State Physics (LASSP),
Clark Hall, Cornell University, Ithaca, New York 14853-2501, USA}


\begin{abstract}
We provide a minimal continuum model for mesoscale
plasticity, explaining the cellular dislocation structures observed in
deformed crystals. Our dislocation density tensor evolves from random,
smooth initial conditions to form self-similar structures strikingly
similar to those seen experimentally -- reproducing both the fractal
morphologies and some features of the scaling of cell sizes and
misorientations analyzed experimentally. Our model provides a framework
for understanding emergent dislocation structures on the mesoscale, a
bridge across a computationally demanding mesoscale gap in the
multiscale modeling program, and a new example of self-similar structure
formation in non-equilibrium systems.
\end{abstract}


\maketitle
\begin{figure}[!t]
\centering
\includegraphics[width=1.0\columnwidth]{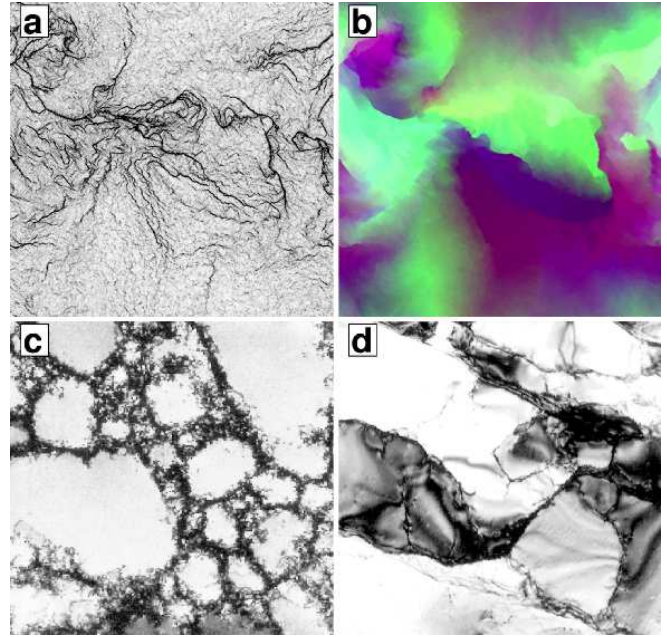}
\caption{{\bf Theoretical and experimental dislocation fractal morphologies}~(color online). 
{\it Top:} Simulated fractal cell wall pattern after 
uniaxial strain of $\epsilon_{zz}=4\beta_0$. (a) Dislocation density plot; (b) Local orientation map. 
{\it Bottom:} TEM micrographs taken from: (c)  
a Cu single crystal~\cite{MughrabiPMA86} after [100] tensile deformation to a stress
of 75.6 MPa
and (d) an Al single crystal following compression to $\epsilon=0.6$~\cite{HughesPRL98}, respectively.
Gray scales have been adjusted to facilitate visual comparisons. Note the striking morphological similarity between
theory and experiment.}
\label{fig:TheoryVSExp}
\end{figure}

Structural engineering materials have a bewildering variety
of microstructures, which are often controlled by deformation and
annealing during the formation process. An imposed distortion 
generates a complex morphology even for a single crystal of a
pure material -- polycrystalline grains form at high temperature where
dislocation climb allows for polygonization, cell structures form at low
temperatures when climb is forbidden. Cell walls (Fig.~\ref{fig:TheoryVSExp}c,d) are distinct from 
grain boundaries in that they have smaller misorientations, different origins,
are morphologically fuzzier, and the cells refine (get smaller) under
shear. Experiments differ in
characterizing the cell structures; some show convincing evidence of
fractality~\cite{MughrabiPMA86,SchwinkSM92,ZaiserPRL98}
with structure on all length scales (Fig.~\ref{fig:TheoryVSExp}c), while others show structures
with a single characteristic scale setting their cell size and cell wall
misorientation distributions~\cite{HughesAM97,HughesPRL98,HughesPRL01}
(Fig.~\ref{fig:TheoryVSExp}d).

Dislocation avalanches~\cite{miguel01Nature}, size-dependent hardness (smaller is
stronger)~\cite{uchic04Science} and cellular structures~\cite{MughrabiPMA86,HughesPRL98} all emerge from collective dislocation
interactions on the micron scale. We expect that these mesoscale phenomena
should be captured by an appropriate continuum theory of dislocation 
dynamics. Computationally, such a theory is crucial for multiscale
modeling, as atomistic and discrete dislocation simulations are challenging
on these scales of length and strain. Here we present a minimal model
for cellular structures, which eventually can be extended to include
the pinning and entanglement needed for avalanches and hardness,
and the slip systems and statistically stored dislocations
needed for realistic descriptions of texture evolution and cross-slip~\cite{mach10JMPS}.
Our model gives the elegant, continuum explanation for the formation and
evolution of cellular dislocation structures. It exhibits both the
experimentally observed fractal structures and scaling collapses
hitherto thought incompatible. Finally, it provides the fundamental
distinction between cell walls and grain boundaries; cell walls are
intrinsically branched in a fractal fashion.

Within a continuum theory of dislocation
dynamics~\cite{AcharyaJMPS05,JSPRL06}, incorporating only elastic
self-interactions with a minimally modified gradient dynamics,
we study the relaxation of a smoothly deformed crystal and
its subsequent evolution under external strain (Fig.~\ref{fig:TheoryVSExp}a and~b).
When climb is allowed, we find the distortion neatly evolves into a stress-free
collection of grain boundaries. When climb is forbidden, cell wall
structures evolve with power-law correlations and self-similarity --
providing a clear morphological distinction between cell walls and grain
boundaries, a tangible model for the experimentally observed
fractal structures~\cite{MughrabiPMA86,SchwinkSM92,ZaiserPRL98},
and an alternative to those that predict microstructure without a 
wide range of scales~\cite{OrtizRS99}.
The resulting morphology, however, is self-similar only after rescaling both
space {\em and amplitude}. Performing the experimentalist's analysis 
of the misorientations and cell size
distributions~\cite{HughesAM97,HughesPRL98,HughesPRL01} yields good agreement with the observed 
scaling form (albeit with
significantly different scaling functions and exponents). By directly exhibiting
key features of the emergent experimental behavior in a continuum, deterministic
dislocation density theory, our simulations pose a challenge to theories
based on stochasticity in the continuum laws~\cite{ZaiserPRL98,GromaPRL00}
or in the splittings and rotations of the macroscopic
cells~\cite{Pantleon98,JSPRB03}. Can these stochastic theories describe our
chaotic dynamics after coarse graining?

Our order parameter is the plastic distortion tensor $\betaP$.
Together with the resulting elastic distortion $\betaE$ derivable from
$\betaP$ via
the long-range fields of the dislocations~\cite{JSPRL06}, $\betaP$
both gives the deformation $\vect{u}$ of the material
(through $\partial_i u_j = \betaE_{ij} + \betaP_{ij}$) and gives a three-index
variant of the Nye dislocation density tensor~\cite{Nye53}
$\rho_{ijk}(\bx) = \partial_j\betaP_{ik}-\partial_i\betaP_{jk}$
(defining the flux of dislocations with Burgers vector along the coordinate
axis $\hvect{e}_k$ through the infinitesimal surface element along
$\hvect{e}_i$ and $\hvect{e}_j$). 
$\betaP$ thus
fully specifies 
the dislocation wall morphologies,
the crystal rotation
(the Rodrigues vector $\vect{\Lambda}$ giving the axis and angle of rotation),
and the stress field $\sigma$ 
(the external load plus the long-range stresses from the
dislocations, given by a kernel~\cite{JSPRL06,TMB91}
$\sigma_{ij}(\mathbf{r}) = \sigma_{ij}^{\mathrm{ext}} 
	+ \int K_{ijkl}(\mathbf{r}-\mathbf{r}') \rho_{kl}(\mathbf{r}')\,d{\mathbf{r}'}$).

\begin{figure}[!t]
\centering
\includegraphics[width=1.0\columnwidth]{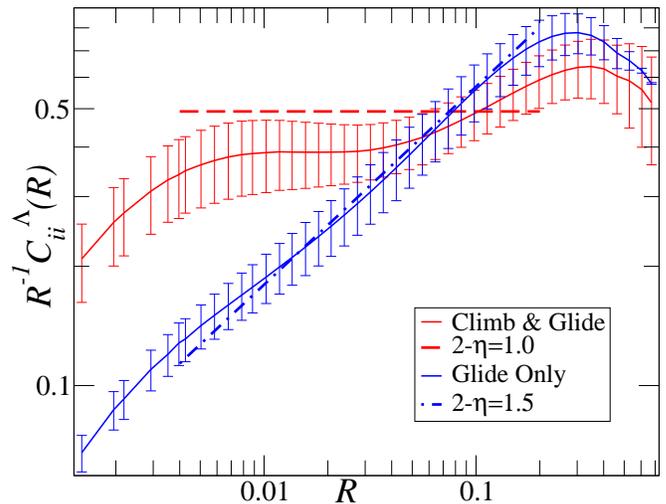}
\caption{ {\bf Scaling of the correlation function} ($1024^2$ simulations).
The trace of the orientation-orientation correlation function
$C^{\Lambda}_{ij}(R) = \langle (\Lambda_{i}(\vect{x})-
			\Lambda_{i}(\vect{x+r}))(\Lambda_{j}(\vect{x})-
			\Lambda_{j}(\vect{x+r}))\rangle$
is averaged over all pairs of points at distance $|\vect{r}|=R$.
Notice that the simulation allowing climb has
$C^{\Lambda}_{ii}(R)\sim R$ as expected for non-fractal grain boundaries. Notice that
the glide-only simulations show $C^{\Lambda}_{ii}(R) \sim R^{2-\eta}$ with
$\eta \approx 0.5$, indicating a fractal, self-similar cell structure,
albeit cut off by lattice and system size effects.}
\label{fig:CorrelationFunctions}
\end{figure}

\begin{figure}[t]
\centering
\includegraphics[width=1.0\columnwidth]{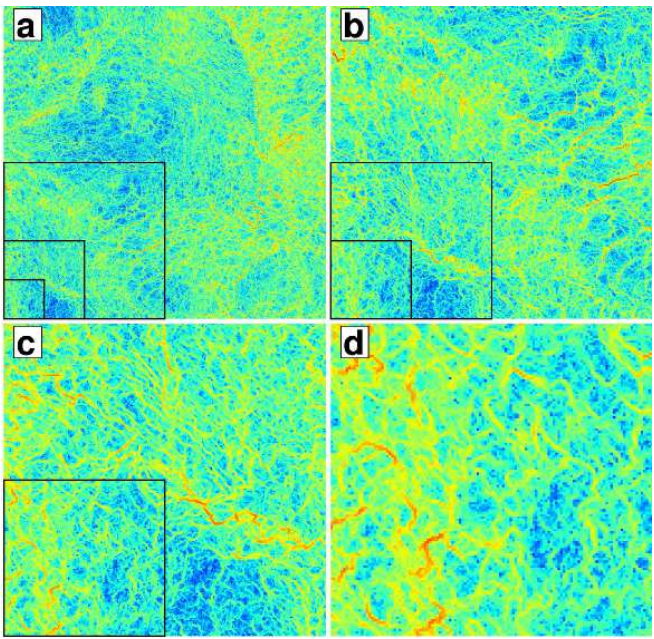}
\caption{{\bf Self-similarity in real space}~(color online). 
Each frame represents the lower left-hand quarter of the previous frame.
Frame (a) is a $1024\times1024$ simulation;
(b), (c), and (d) are thus of length $L=512$, $256$, and $128$.
All are rescaled in amplitude by $(L/L_0)^{-\eta/2}$ with $\eta=0.5$ 
(see Fig.~\ref{fig:CorrelationFunctions} and Table~\ref{table:CorrelationExponents}).
The scale is logarithmic with a range of almost $10^{7}$.
Notice the statistical self-similarity. Other regions, when expanded,
can show larger differences between scales,
reflecting the macroscopic inhomogeneity of the dislocation density.}
\label{fig:RealSpaceRG}
\end{figure}

\begin{figure*}[!t]
\centering
\includegraphics[width=1.27\columnwidth]{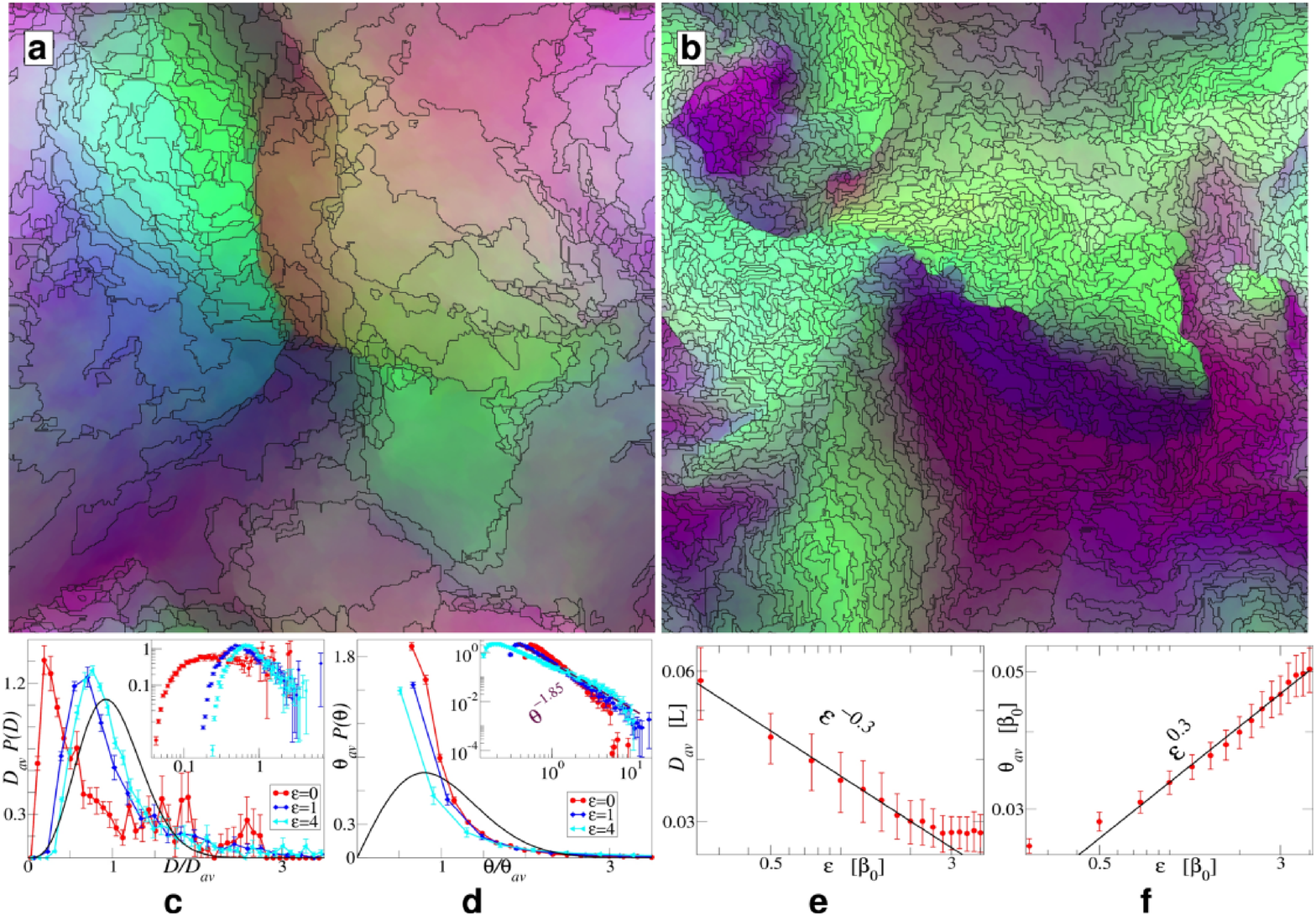}
\caption{ {\bf Cellular structures under strain: size
and misorientation distributions}~(color online). 
(a)~An unstrained state formed by relaxing a random deformation, decomposed
into cells determined by our boundary pruning method: we systematically
remove boundaries in order of their average misorientation angle, and then
prune cells based on their perimeter/area ratio and misorientation angle
(see Supplementary Movie~4). Boundaries below a threshold 
root-mean-square misorientation $\theta_c=0.015 \beta_0$ are removed. 
(b)~The final state after a strain of $\epsilon_{zz} = 4 \beta_0$
is applied; notice the cell refinement to shorter length scales.
(c)~The cell size distribution (square root of area), scaled by the 
mean cell size and weighted by the area, at various external strains. 
(d)~The misorientation 
angle distribution, weighted by cell boundary length, scaled by the mean.
For each curve, data starts at $\theta_c$.
This distribution appears to be closer to a power-law (inset) than
the experimental distributions
(solid curves~\cite{HughesAM97,HughesPRL98,HughesPRL01}). 
(e,f) Mean cell size $D_{av}$ and misorientation angle $\theta_{av}$
as functions of external strain. We find these same power-laws $D_{av}\sim\epsilon^{-0.26\pm0.14}$
and $\theta_{av}\sim\epsilon^{0.26\pm0.04}$, with errors reflecting over a range of
$\theta_c$ and for a variety of pruning algorithms and weighting functions.
Notice that the product $D_{av} \theta_{av}$
is approximately constant, as observed experimentally~\cite{HughesAM97}.
The power-law dependence $\epsilon^{0.3}$ is weaker than the
powers $\epsilon^{1/2}$ and $\epsilon^{2/3}$ observed experimentally
for incidental dislocation boundaries and geometrically necessary boundaries,
respectively.}
\label{fig:Histograms}
\end{figure*}

Following Roy and Acharya~\cite{AcharyaJMPS05},
we assume the flow of $\rho_{ijk}(\bx)$ is characterized by a single velocity
$\vect{v(\bx)}$. Allowing both climb and glide, we can take the velocity $\vect{v}$ to be
proportional to the Peach-Kohler force $\vect{\F}$ on the entire population of dislocations times
a mobility $D(|\rho|)$
$v_a = D(|\rho|) \F_a = D(|\rho|) \rho_{ast} \sigma_{st}$, where $\sigma$ is the stress; 
we then define
$\partial{\betaP}_{ij}/\partial t = J_{ij} = v_a \rho_{aij}$. (This
provides the same equation of motion derived later by Limkumnerd and
Sethna~\cite{JSPRL06}.) To remove dislocation climb (mass transport via
frozen-out vacancy diffusion), we must set the trace of the
volume change $J_{ii}=0$, suggesting a dynamics which moves only the traceless
portion of the dislocation density:
\begin{equation}
\frac{\partial \beta^P_{ij}}{\partial t}=J_{ij}=v_a \rho_{aij}-
\frac{1}{3}\delta_{ij}v_a \rho_{akk}.
\label{eq:Dynamics}
\end{equation}
In this case, to guarantee that energy monotonically decreases we
are led to choose the velocity based on the Peach-Kohler force on this 
traceless part
$v_a=D(|\rho|)(\rho_{ast}-\delta_{st}\rho_{abb}/3)\sigma_{st}$,
making the rate of change of the energy density the negative of a perfect
square~\cite{tobepublished}. (This differs from our 
earlier glide-only formulation~\cite{JSPRL06}.)
To ensure that the velocity is proportional to the force per dislocation,
we choose $D(|\rho|) = 1/|\rho| = 1/\sqrt{\rho_{ijk} \rho_{ijk}/2}$.
Our theory does not incorporate effects of 
disorder, dislocation pinning, entanglement, glide planes,
crystalline anisotropy, or geometrically unnecessary
dislocations. It is designed to provide the simplest framework for
understanding dislocation morphologies on this mesoscale.
\begin{table}
\caption{\label{tab:table4}{\bf Critical exponents} measured for different correlation functions. {\bf GO:} Glide Only;
{\bf CG:} Climb\&Glide; {\bf ST} Scaling Theory~\cite{tobepublished}.}
\begin{ruledtabular}
\begin{tabular}{lcccc}
$\;\;$Correlation functions&GO&CG&ST\\
\hline
$C^{\Lambda}_{ii}(\mathbf{r})\;=\langle\sum_i[\Lambda_i(\mathbf{r})-\Lambda_i(0)]^2\rangle$&$1.5\pm0.1$&$1.1\pm0.1$&$2-\eta$\\
$C^{\rho}(\mathbf{r})\;\,=\langle[\rho_{ij}(0)\rho_{ij}(\mathbf{r})]\rangle$&$0.4\pm0.1$&$0.9\pm0.3$&$\eta$\\
\end{tabular}
\end{ruledtabular}
\label{table:CorrelationExponents}
\end{table}

Our simulations show a close analogy to those of turbulent flows. 
As in three-dimensional turbulence, defect structures mediate 
intermittent transfer of morphology to short length scales. (Unlike
two-dimensional turbulence, we find no evidence of an inverse cascade --
our simulations develop structure only at scales less than or equal to
the initial correlation length of the deformation field.) As
conjectured~\cite{Siggia92PFA} for the infinite-Reynolds
number Euler
equations, our simulations develop singularities in finite time~\cite{JSPRL06}.
It is unclear whether our physically motivated equations
have weak solutions; our simulations exhibit statistical convergence,
but the solutions continue to depend on the lattice cutoff (or on the
magnitude of the artificial diffusion added to remove lattice effects)
in the continuum limit~\cite{tobepublished}. Since our simulations exhibit structure down
to the smallest scales, we conjecture that this is a kind of sensitive
dependence on initial conditions -- but here amplified not by passage of
time, but by passage through length scales. Since the physical system
is cut off by the atomic scale, we may proceed even though our equations
are in some sense unrenormalizable in the ultraviolet.

We simulate systems of spatial extent $L$ in two dimensions with periodic
boundary conditions; our deformations, rotations,
strains, and dislocations are fully three-dimensional.
The initial plastic distortion
field $\betaP$ is a Gaussian random field with decay length $L/5$ and
initial amplitude $\beta_0 = 1$.
We apply a second order central upwind scheme designed for Hamilton-Jacobi
equations~\cite{Kurganov01} on a finite difference grid. The unstrained simulations
presented are at late time, where the elastic energy density is
small and smoothly decreasing to zero, (see Supplementary Movies~1 and~2). 
The strained simulations in Fig.~\ref{fig:Histograms}, (see Supplementary Movie~3),
have uniaxial strain 
in the out-of-plane direction, which is increased
by adjusting the external stress $\sigma_{zz}(t)$ to hold $\epsilon(t)$
fixed. The strain rate is $\dot\epsilon = 0.05 \beta_0^2$. 

Figure~\ref{fig:CorrelationFunctions} shows the
orientation-orientation correlation function. Here we see that
the cellular (climb-free) structures have non-trivial power-law scaling,
but we see non-fractal behavior in the grain boundary morphology
allowing climb. In Supplementary Movie~2, the complex structure of cell walls (climb-free)
shows a few primary large-angle boundaries with high dislocation
density and many low-angle sub-boundaries, leading to fuzzy cell walls
that are qualitatively different from the grain boundaries (climb~\&~glide, seen in 
Supplementary Movie~1). 
Table~\ref{table:CorrelationExponents} includes also 
the correlation function of the
total dislocation density; one can show~\cite{tobepublished}, if the elastic strain is zero~\cite{JSPRB07},
that $C^{\rho}(\mathbf{r}) = -\partial^2 C^{\Lambda}_{ii}(\mathbf{r})-\partial_i\partial_k C^{\Lambda}_{ik}(\mathbf{r})$, 
so $C^{\Lambda}_{ij}(\mathbf{r})\sim |\mathbf{r}|^\alpha$ tells us that $C^{\rho}(\mathbf{r})\sim|\mathbf{r}|^{\alpha-2}$,
implying the exponent relation $\alpha = 2-\eta$ in the last column of Table~\ref{table:CorrelationExponents}.
The scaling for the correlation function for the total plastic distortion $\betaP$ is not as 
convincing~\cite{tobepublished}.
Both are consistent with a renormalization-group transformation that
rescales the dislocation density by a factor of $b^{-\eta/2}$ when it
rescales the length scale by a factor of $b$. Figure~\ref{fig:RealSpaceRG}
gives a real-space renormalization-group illustration of this
self-similarity; the cell walls form a self-similar, hierarchical structure. 

Can we reproduce the experimental fractal characterization of cell boundaries?
Box-counting applied to the dislocation density
(as in Fig.~\ref{fig:TheoryVSExp}a) gives dimensions that depend strongly
on the amplitude cutoff (the dislocation density is self-similar, not
a simple fractal). If we first decompose our simulation into cells 
as in Fig.~\ref{fig:Histograms}b, and apply box-counting to the 
resulting cell boundaries, we obtain a 
fractal dimension of around 1.5 over about a decade~\cite{tobepublished},
compared to the experimental values of $1.64-1.79$~\cite{ZaiserPRL98}.
Such a measurement, however,
ignores the important variation of wall misorientations with scale (capturing
the spatial scaling but missing the amplitude scaling).

Can we reconcile our self-similar cell morphologies with the experimental
analyses of Hughes and collaborators~\cite{HughesAM97,HughesPRL98,HughesPRL01}? Using
our boundary-pruning algorithm to identify cell walls,
Fig.~\ref{fig:Histograms}c and~d show the cell size and misorientation
distributions extracted from an ensemble of initial conditions. The
misorientation distribution we find is clearly more scale free (power-law) than 
that seen experimentally. Under external strain,
we do observe the experimental cell structure refinement
(Fig.~\ref{fig:Histograms}b), and we find the
experimental scaling collapse of the cell-size and misorientation distributions
(Fig.~\ref{fig:Histograms}c and~d) and the observed power-law scaling of the
mean size and angle with external strain (Fig.~\ref{fig:Histograms}e and~f),
albeit with different scaling functions and power-laws than those seen 
in experiments~\cite{HughesAM97,HughesPRL98,HughesPRL01}.

Because we ignore slip systems,
spatial anisotropy, and immobile and geometrically unnecessary dislocations,
we cannot pretend to reflect real materials. But by distilling these
features out of the analysis, we have perhaps elucidated the fundamental
differences between cell walls and grain boundaries, and provided a new example of non-equilibrium scale invariance. 

\begin{acknowledgments}
We would like to thank S. Limkumnerd, P. Dawson, M. Miller, A. Vladimirsky, R. LeVeque,
E. Siggia, and S. Zapperi for helpful and inspiring discussions on
plasticity and numerical methods and N. Hansen, D.C. Chrzan, H. Mughrabi and M. Zaiser for permission of
using their TEM micrographs. We were supported by DOE-BES through
DE-FG02-07ER46393. Our work was partially supported by the National Center for
Supercomputing Applications under MSS090037 and utilized the Lincoln and Abe clusters.
\end{acknowledgments}


\end{document}